\documentclass[prb,showpacs,twocolumn,superscriptaddress]{revtex4-1}

\usepackage{amssymb}
\usepackage{graphicx}
\usepackage{bm}
\usepackage{xcolor}
\usepackage{epsfig}
\usepackage{amsfonts}

\newcommand{\beqa}{\begin{eqnarray}}
\newcommand{\beeq}{\begin{equation}}
\newcommand{\eeqa}{\end{eqnarray}}
\newcommand{\eeqe}{\end{equation}}

\newcommand{\cun}{Hirjibehedin06,Hirjibehedin07,Otte08,Otte09,Loth10a,Loth10b,Loth12,Bryant13,Oberg14,Choi14, Spinelli14,Bryant14,Loth2015,Yan2015,Heinrich2015,Bergmann15,Bryant15}

\newcommand{\model}{JFR09,Fransson,Lorentechains,ZitkoNJP1,Delgado10a,Delgado10b,Delgado14,Sanvito3rd,Ternes15}

\newcommand{\DFT}
{Hirjibehedin07,ZitkoNJP1,Shick_DFT,Rudenko,Lorente09,Nicklas_DFT,Urdainz1_DFT,Yuan_DFT,Pushpa_DFT,Urdainz2_DFT}

\newcommand{\othertheory}
{Persson09,Koenig10,Fransson10,Nicoreview12,Delgado15,Goldberg15}

\newcommand{\theory} {\model,Delgado11,\DFT,\othertheory}

\begin{document}

\title{Electronic properties of transition metal atoms on Cu$_2$N/Cu(100)}

\author{A. Ferr\'on}
\affiliation{International Iberian Nanotechnology Laboratory (INL), Av. Mestre Jos\'e Veiga, 4715-330, Braga, Portugal}

\affiliation{Instituto de Modelado e
Innovaci\'on Tecnol\'ogica (CONICET-UNNE), Avenida Libertad 5400, W3404AAS
Corrientes, Argentina.}

\author{J. L. Lado}

\author{J. Fern\'andez-Rossier}
\altaffiliation{Permanent address: Departamento de F\'{\i}sica Aplicada, 
Universidad de Alicante.}

\affiliation{International Iberian Nanotechnology Laboratory (INL), Av. Mestre Jos\'e Veiga, 4715-330, Braga, Portugal}

\begin{abstract}
We study the nature of spin excitations  of individual transition metal
atoms (Ti, V, Cr, Mn, Fe, Co and Ni) deposited on a Cu$_2$N/Cu($100$) surface using both 
spin-polarized density functional theory (DFT) and exact
diagonalization of an Anderson model derived from DFT.
We  use DFT to compare the structural, electronic and magnetic 
properties of  different transition metal adatoms on the surface. We find 
that the average occupation of the transition metal $d$ shell, main 
contributor to the magnetic moment, is not quantized, in contrast 
 with the quantized spin in the model Hamiltonians that successfully  describe  spin excitations in this system. 
 In order to reconcile these two pictures,  we build a multi-orbital Anderson 
Hamiltonian for the $d$ shell of the transition metal hybridized with the $p$ 
orbitals of the adjacent Nitrogen atoms, by means of maximally localized 
Wannier function representation of the DFT Hamiltonian. The exact solutions of 
this model have quantized total spin, without quantized charge at the $d$ shell.  
We propose that  the  quantized spin of the  models actually  belongs 
to many-body states  with two different charge configurations
 in the $d$ shell,  hybridized  with the $p$ orbital of the adjacent Nitrogen
atoms.  This scenario  implies that the measured  spin excitations are not fully
localized at the transition metal.
\end{abstract}
\date{\today}

\maketitle

\section{Introduction}
The Cu(100) surface coated with a  Cu$_2$N monolayer has turned out to be a 
 remarkable system\cite{\cun} to probe and engineer the electronic properties 
of individual transition metal atoms using scanning tunneling microscopy (STM) and
inelastic electron tunneling spectroscopy (IETS).  
A variety of breakthroughs have been 
reported 
on this system, such as the first measurement of the magnetic anisotropy of 
an individual quantized spin  by means of IETS,\cite{Hirjibehedin07}
the demonstration of single atom spin torque,\cite{Loth10a} the  fabrication 
of nano engineered chains both with   antiferromagnetic\cite{Loth12} and 
ferromagnetic\cite{Spinelli14}  broken symmetry ground states,  probed by means of spin-polarized STM, the 
measurement of spin excitations in spin chains with strong quantum fluctuations that prevent
 spin symmetry breaking,\cite{Hirjibehedin06,Bryant13}  the measurement of single spin relaxation 
time by means of voltage pulse pump-probe technique,\cite{Loth10b}  the observation of renormalization of
magnetic anisotropy due to Kondo exchange interactions\cite{Oberg14} and  the imaging of spin wave modes with atomic scale resolution.\cite{Spinelli14}

The system has been studied from the theoretical standpoint, using a variety of approaches.\cite{\theory}
Importantly, both the  spin excitation spectra of individual 
atoms\cite{Hirjibehedin07,Otte08,Yan2015} and  multi-atom structures,
\cite{Hirjibehedin06,Otte09,Loth12,Bryant13,Oberg14,Spinelli14,Bryant14,
Loth2015} as well as their spin relaxation dynamics have been successfully described using model Hamiltonians\cite{\model}  where  {\em quantized} spins interact with each other via Heisenberg coupling,\cite{Hirjibehedin06, JFR09, Lorentechains,Spinelli14} and are Kondo coupled both to the tunneling electrons\cite{JFR09,Fransson} and to the substrate.\cite{Delgado10a,Delgado10b,Oberg14,Delgado14} 
Treating   Kondo coupling up to second order in perturbation theory accounts for spin relaxation\cite{Delgado10a,Delgado10b} and magnetic anisotropy renormalization.\cite{Oberg14,Delgado14}
Furthermore,  calculations\cite{Sanvito3rd,Ternes15} up to third order are also able to account for non-trivial  IETS line shapes, including Kondo peaks.  Numerical renormalization group 
non-perturbative calculations  for the anisotropic spin Kondo model also provide very 
good description for both the finite energy spin excitations and the zero 
bias Kondo peak in these systems. \cite{ZitkoNJP1,ZitkoNJP2,ZitkoNJP3}
. The origin of the Kondo couplings in these systems can be traced down to a 
multi-orbital
Anderson model for these $S>1/2$ systems,\cite{Delgado11}  in line with the very well known results 
for the mapping\cite{Schrieffer_old,Anderson1_old, Appelbaum} of the single orbital
Anderson model\cite{Anderson2_old} to the Kondo model.

In spite of the success of quantized spin  Hamiltonians to describe many experiments of transition metals on Cu$_2$N, there is a
  problem of principle that we address in this paper.  Density functional theory (DFT) 
calculations,   most of them\cite{\DFT} dealing with  Ti, Mn, Fe and Co adatoms on Cu$_2$N/Cu(100),  show that nor the charge, neither the magnetic moment of these magnetic atoms are quantized.

Here we  provide a comprehensive and comparative study of the 
electronic and structural properties of the entire series of $3d$ transition metals (Ti, V, Cr, Mn, Fe, Co and Ni).
To the best of our knowledge, no DFT calculations have been reported for 
 V and Ni on Cu$_2$N/Cu(100).  Our calculations confirm the fractional nature of the average occupation of the $d$ levels in these systems,   which 
 is not surprising given  their conducting nature,  but it poses 
an apparent contradiction with the quantized spin model description.  In the second part of the manuscript
we provide a solution to this apparent conflict. We build a multi-orbital Anderson model,  using as starting point the representation of the DFT Kohn-Sham Hamiltonian in a basis of maximally localized Wannier functions.  The Anderson model includes spin-orbit interactions, crystal field interactions, on-site Coulomb repulsion and hybridization of the $d$ shell of the transition metal with its  Nitrogen neighbors. 
  We solve the model by exact numerical diagonalization  within a restricted 
Hilbert space that includes both $d^n$ configurations with $d^{n+1} p^m$ 
configurations, where $n$ stands for the number of electrons in the $d$ shell 
and $m$ stands for the number of electrons in the Nitrogen $p$ orbitals.  Our numerics show  that the low energy excitations of the model can be mapped into quantized spin Hamiltonians.  Within this picture, it is apparent that this quantized spin $S$ describes the quantum number of many-body wave functions that mix states with $n$ and $n+1$ electrons in the $d$ shell, and have thereby  non-integer average occupation. 
  
The paper is organized as follows. In Sec. \ref{sdft}  we describe  DFT 
calculations for different TM atoms at Cu$_2$N paying particular attention
to the structural properties (\ref{st}), electronic properties (\ref{ep}) and
magnetic properties (\ref{smp}). In Sec. \ref{ci} we build the multi-orbital 
Anderson model,  using as starting point the DFT calculations and we analyze
the connection with the Spin Models. Finally,
Sec. \ref{c} contains a summary and a discussion 
our most important findings.


\section{Density Functional Calculations \label{sdft}}

\subsection{Methods\label{cd}}

Most of our DFT calculations of $3d$ transition metal adatoms  adsorbed on  Cu$_2$N/Cu(100) were done
 using the generalized-gradient 
approximation (GGA)  and GGA$+$U for exchange-correlation energy, \cite{gga} 
using 
plane-wave basis sets and Projector Augmented-Wave
(PAW) \cite{paw} as implemented in QUANTUM ESPRESSO (QE) code. \cite{qe}
Additionally, in
some particular cases we have performed complementary calculations using
Local Spin Density Approximation (LSDA) \cite{lda} for exchange-correlation 
energy, using all-electron full-potential linearised augmented-plane wave 
(FP-LAPW) as implemented in ELK. \cite{elk}

\begin{figure}
\begin{center}
\includegraphics[width=0.45\textwidth]{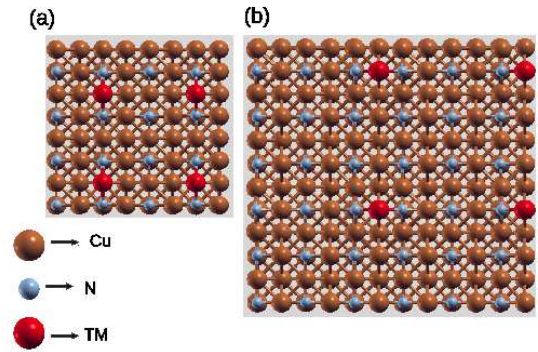}
\end{center}
\caption{(Color online) Structure of the TM$@$Cu$_2$N.  (a)  Small   
supercell where  $a$, the  intercell distance between TM atoms along the N 
direction, is 7.2$\AA$. (b) Big supercell, with $a=10.8 \,\AA$. 
 }
\label{fst}
\label{st1}
\end{figure}

In order to test the size convergence, we have used two super cells with different sizes. 
Both cells have 4 slabs of Cu(100),  separated 
by a vacuum region of $15\,\AA$.  The smaller cell has   $37$ atoms ($1$ TM, $4$ N 
and $32$ Cu ) and a bigger supercell, has  $82$ atoms ($1$ 
TM, $9$ N and $72$ Cu ). The corresponding structures are 
shown in  Fig. \ref{fst}.  In the smallest structure, the  intercell distance 
between TM atoms along the N direction is 7.2$\AA$, while in the bigger one 
this distance is 10.8$\AA$.

QE calculations are done in two stages, structural relaxation and electronic structure calculation.
In the relaxation stage, the mesh in $k$ space for the small and big cells were $6\times6\times1$ and $4\times4\times1$ respectively. 
 In both cases
the relaxation was performed until the forces acting on atoms were smaller
than $10^{-3}\,a.u.$. 
In the second stage,  the meshes used were 
$8\times8\times1$ for the small supercell and $6\times6\times1$ for the big 
one.  In all the calculations we used a smearing with a 
broadening parameter of $0.01-0.02$  $Ry$, in line with previous work,
\cite{Rudenko, Nicklas_DFT,ZitkoNJP1} and we fixed the cutoff energies for 
the wave function and charge density at 40-80 Ry and 400-800 Ry respectively.
For the Elk calculations, we started from the relaxed 
structures obtained with QE. In this case the mesh in 
 $k$ space was  $4\times4\times1$, the  product of the muffin-tin radius and the momentum cutoff is
  $R_{MT}k_{\rm max}=6$ and we employed
Local Spin Density Approximation (LSDA) \cite{lda} for exchange-correlation and
DFT$+$U with Yukawa screening. \cite{yuk}

In the case of Fe and Co we have also obtained the so 
so called maximally localized Wannier functions (MLWF) 
.\cite{wan1,wan2,wan3,wan4,wan5}  associated to the Bloch states of the DFT
calculation, using the package Wannier90. The Wannier functions form an
orthogonal and complete basis set that we can use to describe our system. Importantly, the representation of the
Kohn-Sham Hamiltonian in the DFT basis provides an effective tight-binding model to describe the electronic states of the system, that we use as a starting point to build
a multi-orbital Anderson model, as described in Sec. \ref{ci}.

\subsection{Structural properties  \label{st}}
We now discuss the structural properties of a single $3d$ TM atom bonded to 
the Cu  site of the Cu$_2$N/Cu(100) surface.  This is the binding site most 
frequently reported in the literature.\cite{Hirjibehedin07,Bryant13,Choi14,
Rudenko,Nicklas_DFT,Urdainz1_DFT,Yuan_DFT}  In order to refer to the 
different TM atoms we shall use indistinctly  their chemical formula 
(Ti, V, Cr, Mn, Fe, Co, Ni) or the nominal charge on the $d$ shell 
$q_d=2,3,4,5,6,7,8$. 

\begin{figure}
\begin{center}
\includegraphics[width=0.4\textwidth]{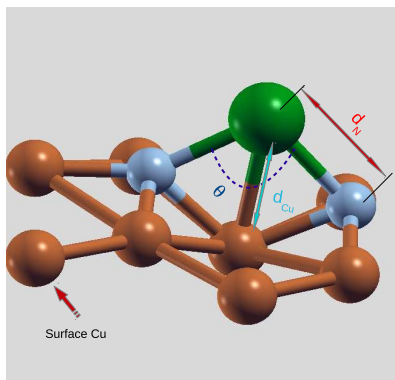}
\end{center}
\caption{(Color online) Schematic diagram of the Cu$_2$N (
Cu atoms in brown,   N atoms in blue) surface with the TM atom (green).
$d_{N}$ is the distance from the TM atom to the nearest neighbors N atoms,
$d_{Cu}$ refers to the distance from the TM atom to the Cu atom lying just 
below the magnetic atom, $\theta$ is the angle formed by the N-TM-N trimer 
and the red arrow shows the position of the Cu atom used to define the 
surface.
 }
\label{car}
\end{figure}

Cu$_2$N is known\cite{Hirjibehedin07,Yuan_DFT} to  form a weakly buckled square
lattice on top of the Cu(100), consistent with our DFT calculations. 
In all cases considered, the adsorption of the TM introduces a local 
distortion on the  Cu$_2$N lattice, shown in Fig. \ref{car}: the underneath 
Cu atom is pushed towards the bulk, and the N atoms are pulled out.  These 
results are in line with previous works.\cite{Yuan_DFT,ZitkoNJP1,gauyacq2010}
In order to characterize this structural distortion, we introduce 3 
distances: the TM- N distance ($d_{N}$), the TM-Cu distance  ($d_{\rm Cu}$) and 
the TM-surface  vertical displacement, $z$, that we take as the $z$ 
component of   the vector that joins the TM  with the  farthest Cu surface 
atom, marked with an arrow in Fig.\ref{car}.  In addition we also introduce 
the angle  formed by the N-TM-N trimer ($\theta$).

Our calculations, performed both for the small and large supercell 
(see Fig. \ref{fst})  show how these structural values are similar for 
different TM atoms, but with  clear and systematic variations as a 
function of the number of $d$ electrons.  Whereas both the TM-N and the TM-Cu 
distances  undergo minor variations across the TM series,   the $\theta$ 
angle has a much more marked change, going from structures where the TM is 
clearly a protrusion and the N atoms are weakly detached from surface, for  
small $q_d$, to structures where TM is almost collinear with the  N atoms, 
for Co and Ni. The tendency to form collinear N-TM-N structures is particularly
clear in the case of Co and Ni chains (small cells).  In the case of Co, 
the marked difference between the small and big cells  suggests that there is 
a  cooperative distortion in the case of Cobalt chains along the N 
direction, \cite{Bryant15}  also visible for Ni,  and clearly absent in
 the case of the lighter TM, such as Fe, for which these chains have been 
studied experimentally.\cite{Loth12,Bryant13}

\begin{figure}
\includegraphics[width=0.5\textwidth]{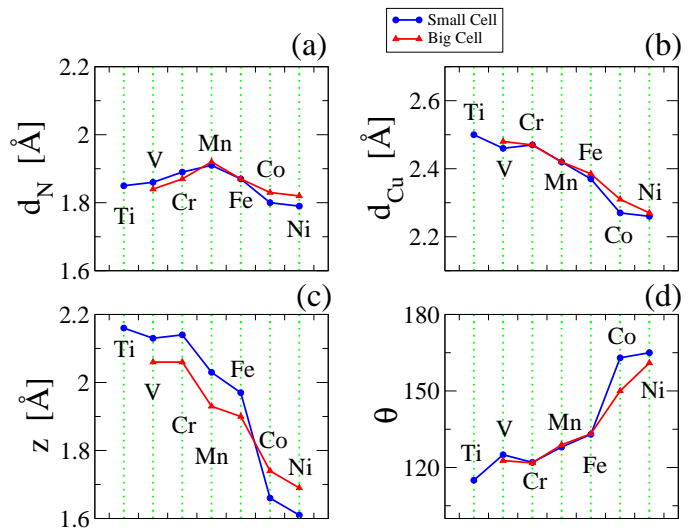}
\caption{(Color online) Comparative study of structural properties of 
TM atoms on Cu$_2$N. (a) Distance form the TM atom to the nearest neighbors 
N atoms $d_{N}$ as a function of the TM atom. 
(b) Distance from the TM atom 
to the Cu atom lying just below the magnetic atom ($d_{Cu}$)  . 
(c) Distance from the TM atom to the Cu$_2$N 
surface ($z$) as a function of the TM atom. 
(d)  Angle formed by the N-TM-N 
trimer.   Blue (red) line shows the result with 
the small (big) cell.
}
\label{ang}
\end{figure}

\subsection{Electronic Properties \label{ep}}

We now discuss the electronic properties of the adsorbed TM on the Cu site of 
Cu$_2$N/Cu(100). Our calculations for the pristine surface show that  the Cu 
and N atoms in the  Cu$_2$N layer have charge  $q_{\rm Cu}=+0.2$ and 
$q_{\rm N}=-0.4$.   The adsorption of the TM atom results in a charge 
transfer  mostly from the  TM $s$ orbitals to the   N ligands, increasing
their negative charge.  The  TM atoms lose practically all the $4s$-electrons. 
The outermost electrons are thereby in the $d$ shell.

\begin{figure}
\begin{center}
\includegraphics[width=0.44\textwidth]{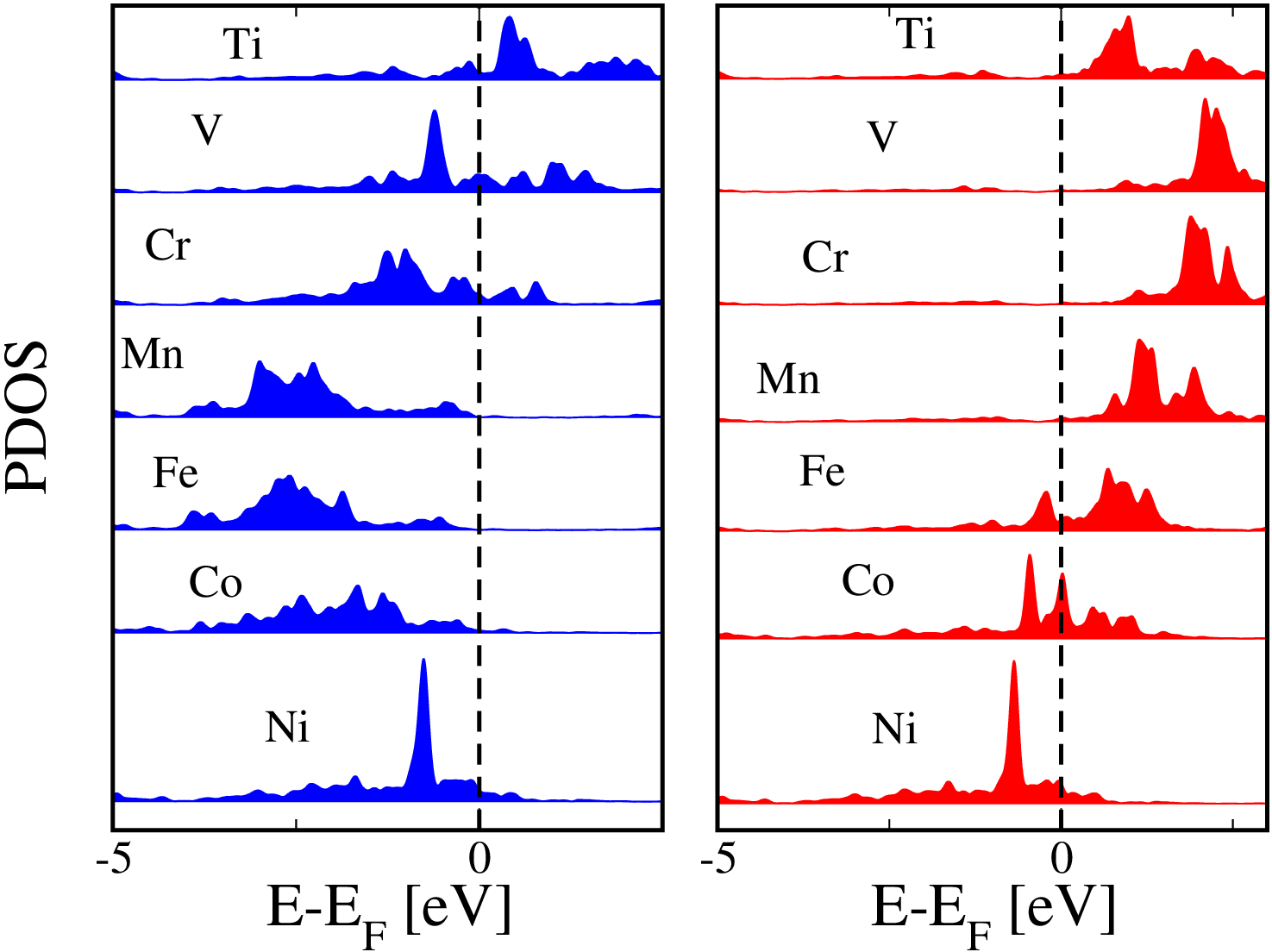}
\end{center}
\vspace{-15pt}
\caption{(Color online) (a)  DOS projected (PDOS) over $d$-orbitals of the
different magnetic atoms, for  majority spin (left panel) and minority spin (right panel).}
\label{dosd}
\end{figure}

A naive  interpretation of the picture that arises from the use of quantized 
spin models to describe these systems  would lead to conclude that the charge 
in the $d$ shell of the TM is quantized.  Our calculations show that this is 
not the case. A hint of this can be already seen by inspection of the 
spin-resolved  density of states projected over the $d$ orbitals of the 
adsorbed TM atoms, shown in Fig. \ref{dosd}, obtained with QE, for the 
small cell.   For instance, the occupancy of the majority spins (left-panel) 
is not 5 for Ni and Co.  The presence of very broad peaks indicates strong 
hybridization of some of the $d$ orbitals with the rest of the system, as we 
show below.    

\begin{figure}
\begin{center}
\includegraphics[width=0.35\textwidth]{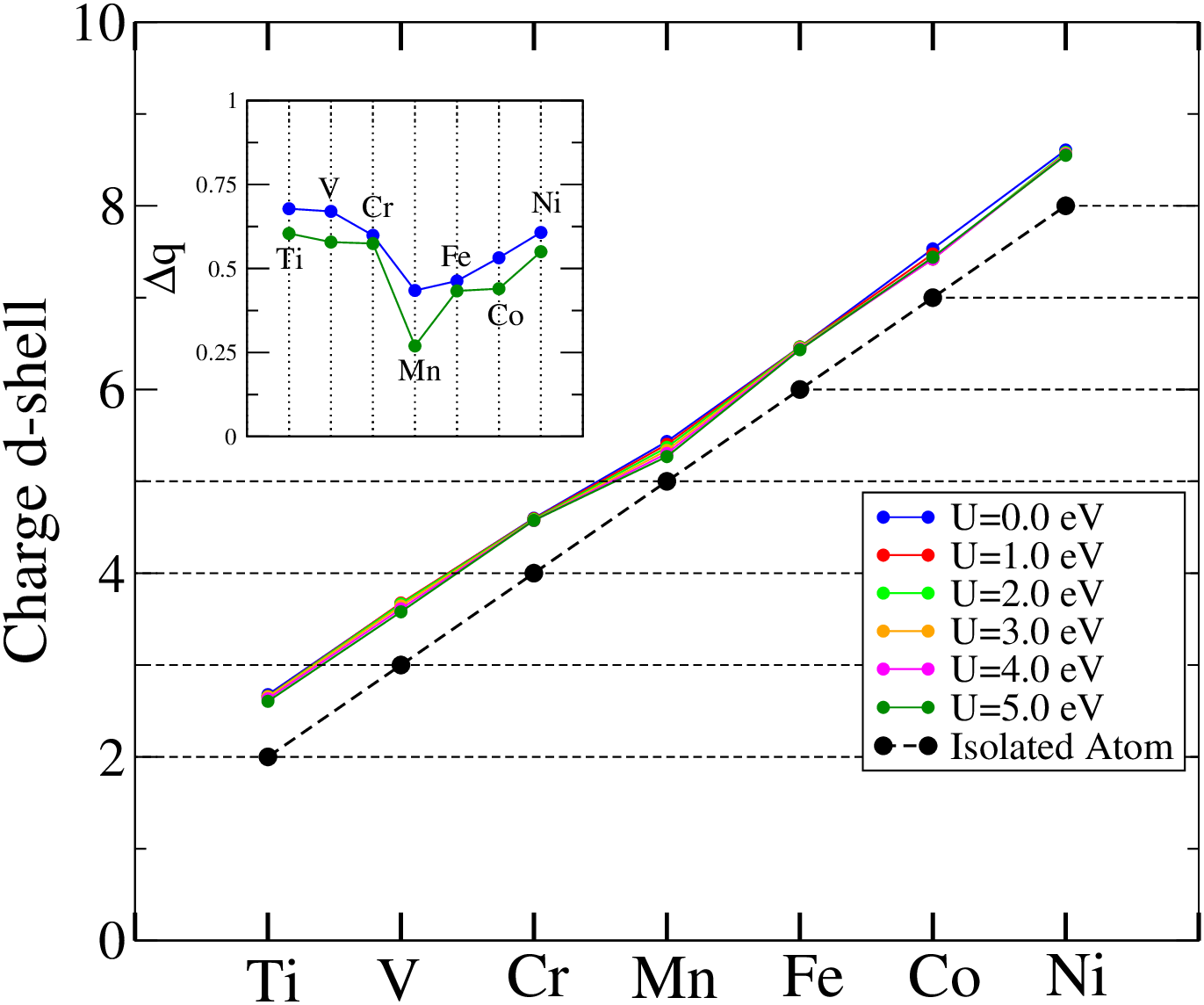}
\end{center}
\caption{(Color online) GGA+$U$  calculation of the total charge
of the $3d$-levels for the different TM atoms and different values of $U$. 
Black dashed line show the result for the isolated atoms. Inset: deviation of 
the charge in the $d$ levels  with respect to the isolated atom}\label{cdu}
\end{figure}

In Fig. \ref{cdu} we show the integrated density of states up to the Fermi 
energy, that gives the  occupation of the $d$-shell
for the different TM atoms, using both GGA and GGA$+U$, for several values 
of $U$.  The results are plotted together with the charge of the isolated atom.
In all instances we find that the $d$ shell is  more charged than in the free 
atom case.

The difference between the computed charge and isolated charge, $\Delta q$, 
is shown in the inset for $U=0$ and $U=5$eV.  Expectedly,  increasing $U$ 
reduces $\Delta q$.  Interestingly, $\Delta q$ increases as we move away 
from half filling (the Mn atom).  Varying $U$ does not yield large changes 
in these results, except for the Mn atom, in line with results
obtained in reference \onlinecite{Yuan_DFT} for Co and Mn.  We have also 
verified that $\Delta q$ is stable with respect to changes in the size of the 
supercell.

\subsection{Magnetic Properties \label{smp}}

\begin{figure}
\includegraphics[width=0.45\textwidth]{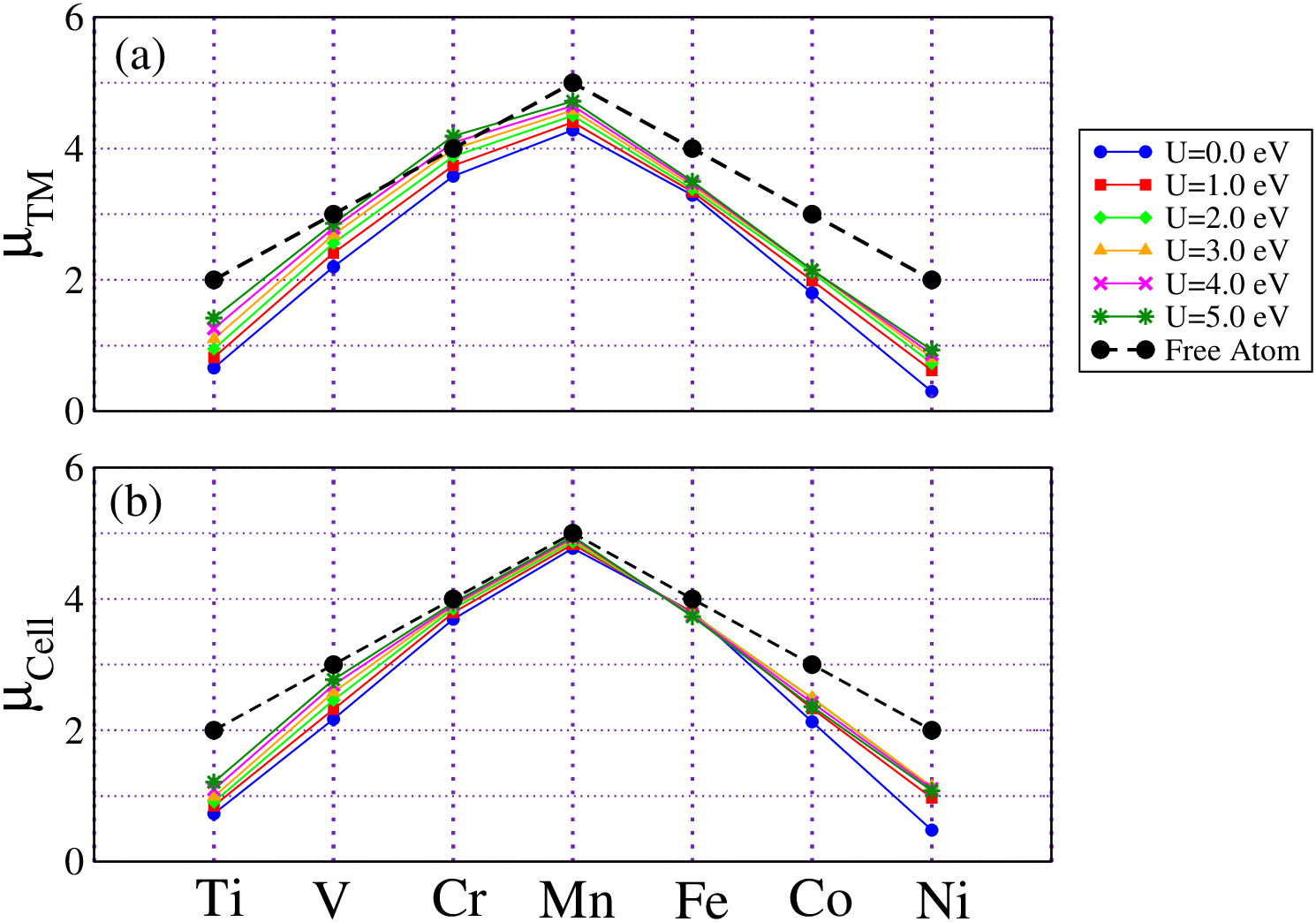}
\caption{(Color online) GGA+$U$ calculation of magnetic moment of the TM atoms (in units of $\mu_B$), as a function of $U$.
 (a) Atomic  (b) Unit cell.  In both panels
the black dashed line shows
the expected magnetic moment for the free (isolated) atom ($\mu_{\rm free}$).
 }
\label{mtm}
\end{figure}

We now discuss the evolution of the magnetic moments,  $\mu$,  of the 
series of 3d  TM  adsorbed atoms. In Fig. \ref{mtm},   we show the magnetic 
moment of the free atoms,  $\mu_{\rm free}$,  as given by   Hund's. The
largest  free atom moment is $\mu=5\mu_B$, for   the half-filled shell (Mn) 
and goes down as we move away from half filling. The upper panel of 
Fig \ref{mtm} shows the GGA$+$U calculation of the magnetic moment of the TM 
atoms. With the only exception of Cr, the magnetic moment for the  
adsorbed TM is always smaller than the free atom case. The deviations become 
particularly severe as we move away from half filling.  For instance, 
the magnetic moment of adsorbed Ni is half the value of the free atom case.  
In contrast with the case of $\Delta q$, the  value of the magnetic moment 
depends more strongly on the value of $U$, yet, the atomic limit is only 
reached in the case of Cr for $U=5$eV.

Whereas most of the spin is localized on the magnetic atoms,  a  substantial 
amount of the spin density is not located at the $d$-levels of the atom.  
Therefore,  we  also plot the cell magnetic moment  and compared them with 
the free atom case (see Fig. \ref{mtm}b). It is apparent that, in the case of the cell moment, 
the deviations from the free case are very small for V, Cr, Mn and Fe. In 
the case of Ti and Ni, the deviation from the free case is still a factor 
of two. The overall trend is that, close to half-filling, the DFT results 
are closer to the free case.

\begin{figure}
\begin{center}
\includegraphics[width=0.8\columnwidth]{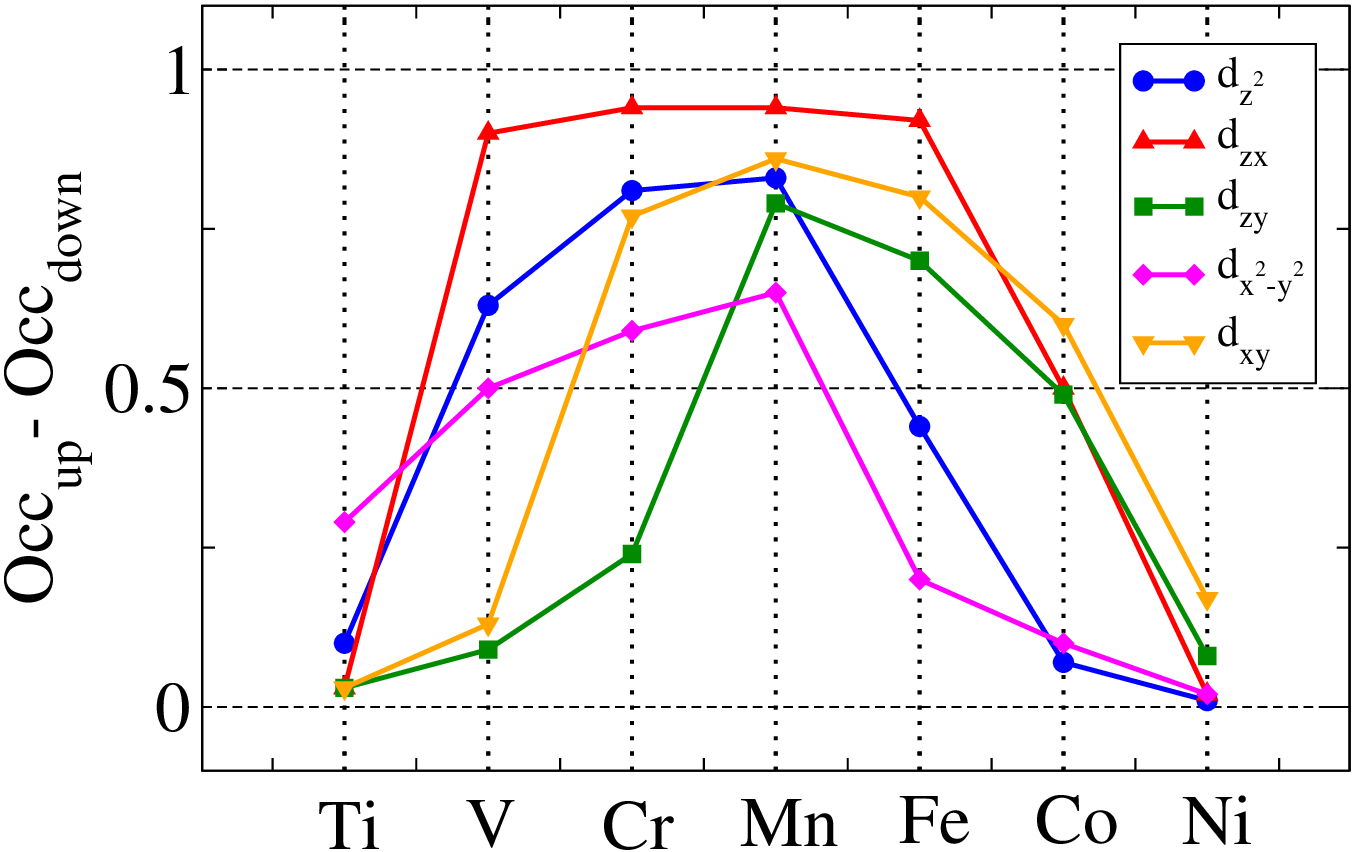}

\end{center}
\vspace{-15pt}
\caption{Orbital breakdown of the spin polarization, calculated using the
small cell and $U=0$.}
\label{mp}
\end{figure}

Fig. \ref{mp} shows the Spin polarization of the different $d$ orbitals as a 
function of the TM atom.  Whereas at half filling the magnetic moment has to 
be evenly distributed in the 5 $d$ orbitals,  away
from half filling this is no longer the case.   The orbital composition of the 
magnetization is interesting because it affects the magnetic anisotropy and 
because only one of the $d$ orbitals, the  $d_z^2$, couples to the  $s$ orbital 
of the last atom in STM tip. This results particularly interesting in the 
case of Ti, Co and Ni where we can appreciate that the $d_z^2$ orbital has a 
very small spin polarization. These calculations were performed for the small
cell, then we should expect interesting features in STM experiments when
dealing with Co and Ni chains. \cite{Bryant15}

\begin{figure}
\includegraphics[width=0.75\columnwidth]{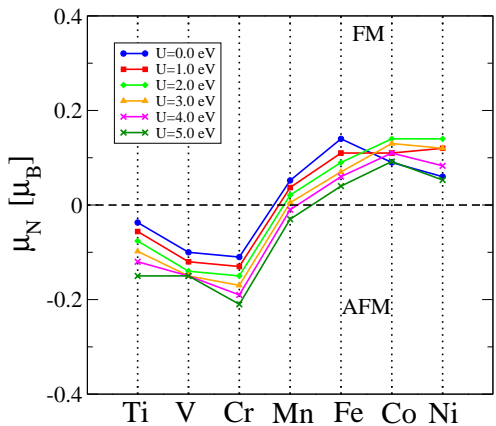}
\caption{(Color online) GGA+$U$ calculation of the
magnetic moment of the nearest neighbor N
atom as a function of the TM atom, for different values of U.}
\label{mn}
\end{figure}

The difference between the atomic and the cell magnetization, summarized in 
Table \ref{t1},  implies that the surrounding atoms gain some magnetic 
moment as well. Our results, shown in Fig. \ref{mn}  also show that their 
alignment with the N atoms can be both ferromagnetic (FM) or antiferromagnetic 
(AFM) depending on the TM. In particular, the correlation is AFM below 
half-filling and FM above half-filling.  At half filling the results depend 
on the value of $U$.  Below we provide an explanation to this sign, 
based on a multi-orbital Anderson model.

\begin {table}
\begin{tabular}{|p{1.1cm}||p{1.15cm}|p{1.4cm}|p{1.15cm}|p{1.45cm}|}
 \hline
 \multicolumn{5}{|c|}{Spin  polarization of TM atoms} \\
 \hline
 Atom& $d$-shell&TM atom&Cell& Free atom\\
 \hline
 Ti &0.62&0.71&0.6 & 1\\
 V  &1.32&1.45&1.41 & 1.5\\
 Cr &1.98&2.10&1.97& 2\\
 Mn &2.25&2.36&2.47& 2.5\\
 Fe &1.62&1.75&1.91& 2\\
 Co &1.05&1.1&1.25& 1.5\\
 Ni &0.46&0.47 &0.57 & 1\\
 \hline
\end{tabular}
\caption {DFT calculation of the spin polarization for  ($U=5$ eV).}\label{t1}
\end {table}

Finally, calculations for Co atoms were performed using ELK for different values
of $U$ in order to check the magnetic properties obtained with QE. The
results obtaining with both codes were in very good agreement.

\section{Connection with the spin models\label{ci}}

The DFT results of the previous section clearly show that the charge and spin of the $d$ shells are not quantized. There is
thus an apparent  conflict between the DFT calculations and the use of spin Hamiltonian models with quantized spins.\cite{Hirjibehedin06,Hirjibehedin07,JFR09} 
In this section we address this important topic and provide a solution for this  conundrum.  First,  we build a multi-orbital Anderson model 
starting from the DFT results. The Anderson model describes the $d$ orbitals of the TM atom, their hybridization to their neighbors, the crystal field splitting due to electrostatic interactions,  intra-shell Coulomb repulsion, and the spin-orbit coupling.  Most of these parameters are obtained  from the DFT calculations, as we discuss below.     Second,   we solve the Anderson model  exactly within a restricted multi-particle Hilbert space that includes both 
 $d^{n}p^{12}$ and $d^{n+1}  p^{11}$
configurations,  where $d^n$ stands for $n$ electrons in 
the  $d$ shell of the TM and the $p^m$ stands for  $m$ electrons in the $p$  
shells of the  Nitrogen first neighbors (without charge transfer
there are 6 electrons in each $p$-shell of the N atoms).   Charge fluctuations 
were shown to be important in the case of Cobalt adatoms on MgO/Ag.\cite{Rau14}
Including charge fluctuations in the Anderson model   gives 
rise to a non-integer occupation of the $d$ shell but still preserves many-body
states with a quantized spin $S$, that is identified with the spin of the 
quantum spin Hamiltonians.

\subsection{Maximally localized  Wannier functions as atomic like basis set
\label{mlwf}}
The  derivation of an effective Anderson model starting from the DFT calculations requires a representation of the Khon-Sham Hamiltonian in a basis set that contains atomic-like $d$ states localized around the transition metal atom.   
Our DFT calculations are performed with  a plane-wave basis
whereas the multiorbital Anderson Hamiltonian demands a local basis.  
Thus, to go from plane wave to a local basis,
 we represent the DFT Kohn-Sham Hamiltonian in the basis  of maximally localized Wannier functions (MLWF),\cite{wan1,wan2,wan3,wan4,lorente-wan1} computed using the code Wannier90, as described  in reference 
\onlinecite{wan4}.  
 
\begin{figure}
\includegraphics[width=0.5\textwidth]{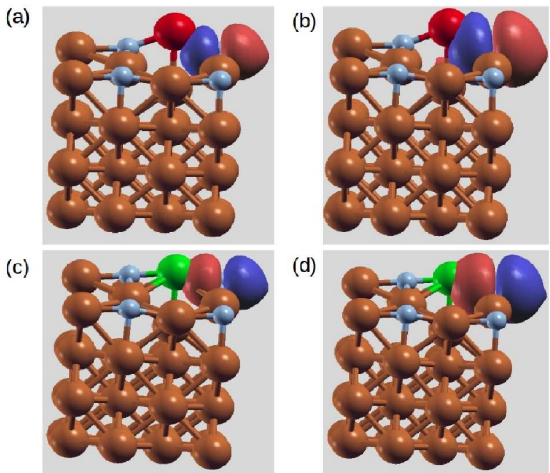}
\caption{(Color online) Wannier orbitals for Fe and Co atoms at Cu$_2$N.
(a)  $p_y$-orbital of a nearest neighbor N atom for Fe at Cu$_2$N close to
the atom (isosurface $3$). (b)  $p_y$-orbital of a nearest neighbor N atom
for Fe at Cu$_2$N far from the atom (isosurface $1$).
(c)  $p_y$-orbital of a nearest neighbor N atom for Co at Cu$_2$N close to
the atom (isosurface $3$). (d)  $p_y$-orbital of a nearest neighbor N atom
for Co at Cu$_2$N far from the atom (isosurface $1$).
 }
\label{wan-1}
\end{figure}

The computation of the MLWF  is implemented as follows. First, we select a group of Bloch
bands from a spin unpolarized \cite{Ferron15} calculation for a given 
TM/Cu$_2$N system.  
An energy window of $16$eV around the Fermi energy
is taken, and the band disentanglement procedure is performed.
The selected  Bloch bands are initially  projected over the $s$, $p$ and $d$ 
orbitals of both the TM atom and the copper atoms  and over the $p$ and $s$ 
orbitals of the nitrogen atoms.   The total number of states involved in this 
procedure is 313, 
corresponding to the 4  Nitrogen atoms (4 orbitals each), 
32 Copper atoms and the TM atom  (9 orbitals each).

\begin{figure}
\includegraphics[width=0.5\textwidth]{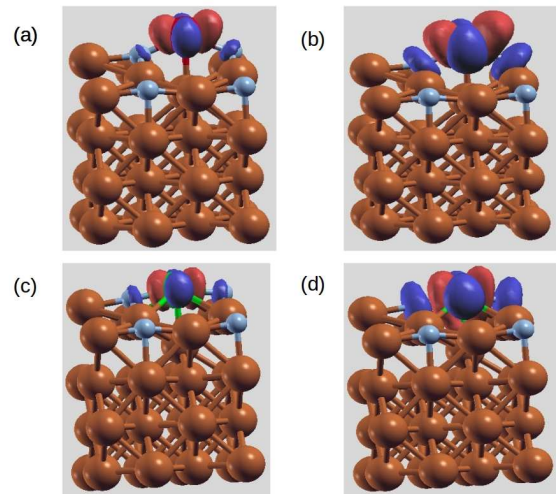}
\caption{(Color online)
Wannier orbitals for Fe and Co atoms at Cu$_2$N.
(a)  $d_{x^2-y^2}$-orbital of a nearest neighbor N atom for Fe at Cu$_2$N
close to the atom (isosurface $3$). (b)  $d_{x^2-y^2}$-orbital of a nearest
neighbor N atom for Fe at Cu$_2$N far from the atom (isosurface $1$).
(c)  $d_{x^2-y^2}$-orbital of a nearest neighbor N atom for Co at Cu$_2$N close
to the atom (isosurface $3$). (d)  $d_{x^2-y^2}$-orbital of a nearest neighbor
N atom for Co at Cu$_2$N far from the atom (isosurface $1$).
 }
\label{wan-2}
\end{figure}

 An iterative procedure   yields  a total of 313 MLWF including 3 $p$-like 
and 5 $d$-like  MLWFs localized around the Nitrogen atoms and the TM atom  
respectively.  They are  shown in Fig. \ref{wan-1} and Fig. \ref{wan-2}, where 
two different iso-surfaces of the MLWF with $y$ and $x^2-y^2$, for Fe and Co, 
are represented.
In the case of iso-surfaces with larger value,  corresponding to the wave-function 
close to  atomic cores, these MLWF have the same symmetry as the
real spherical harmonics with  $L =1$ and $L= 2$. In comparison, 
the iso-surfaces with small value
do not show the symmetry 
of the Cartesian atomic orbitals.  
 
 The Anderson model is build in a basis of single particle states that involves the 5 $d$ orbitals of the TM and the 6 $p$ orbitals of the two first neighbor nitrogen atoms.  The representation of the Khon-Sham Hamiltonian $H_{KS}$  in this basis can be written as: 
 \begin{eqnarray}
 H_{CF}+H_{\rm hyb}= 
 \left(
 \begin{array}{cc}
 H_{dd} & H_{dp}\\
 H_{pd} & H_{pp}
 \end{array}
 \right)
 \end{eqnarray}
where we identify the crystal field  Hamiltonian for the $d$ orbitals  
$H_{CF}=H_{dd}$ and the $pd$ hybridization Hamiltonian with the off-diagonal 
blocks $H_{dp}$ and $H_{pd}$.  Importantly, Fig. \ref{wan-1} and Fig. 
\ref{wan-2} show  how the  MLWF  for Cobalt are more extended than those of Fe,  both for the  $x^2-y^2$ and $y$ orbitals. 
This accounts for the fact that  the hybridization between the TM $d$ orbitals and Nitrogen $p$ orbitals is larger for Co than for Fe.  In particular, the matrix element 
$\langle x^2-y^2|H_{KS}| y\rangle$ is twice as large for Co than for Fe.

\subsection{Multi-orbital  Anderson model\label{gam}}
We now introduce the multi-orbital Anderson model for the TM on the Cu$_2$N substrate describing the electrons in the 5 $d$ orbitals
hybridized with the  $p$ orbitals of the two adjacent $N$ atoms.
The Hamiltonian  is the sum of  4 terms, the Kohn-Sham Hamiltonian of the previous section, plus the  Coulomb repulsion of the electrons
in the $d$ shell, the Coulomb attraction of the $d$ levels with the TM nucleus,  and their spin-orbit coupling:
\beqa
H= H_{KS}  +   H_{{\rm Coul}}+ H_{Z}+H_{\rm SO}
\label{Htot}
\eeqa

In the following we label the  five $d$-like MLWF of the TM with  the index $m$,   and 6 $p$-like MLWF of the two Nitrogen atoms with the index $n$. 
The second quantization representation of the first term reads: 
\begin{eqnarray}
H_{\rm KS} + H_Z&=&  \sum_{m,m',\sigma} \left( \langle m|H_{\rm KS}|m'\rangle  + E_d \delta_{m,m'}\right) d_{m\sigma}^\dag d_{m'\sigma} + \nonumber \\
&+&
\sum_{n,n',\sigma} \langle n|H_{\rm KS}|n'\rangle p_{n\sigma}^\dag p_{n'\sigma} + \nonumber\\ &+&
\sum_{m, n,\sigma} \left(\langle m|H_{\rm KS}|n\rangle d_{m\sigma}^\dag p_{n \sigma} + {\rm h. c.}\right)
\end{eqnarray}
The matrix $H_{dd}= \sum_{m,m',\sigma} \langle m|H_{\rm KS}|m'\rangle$ describes the crystal field. \cite{Ferron15}
The $E_d$ energy scale  accounts for the Coulomb interaction with the positive charge in the nucleus and has to be included to offset the excess in 
Coulomb repulsion in the configurations $d^{n+1}$  with an  extra electron.  The matrix $H_{pp}=\langle n|H_{\rm KS}|n'\rangle $
 describes the single-particle $p$ levels of the first neighbor Nitrogen atoms and $H_{dp}=\langle m|H_{\rm KS}|n\rangle$ describes their hybridization
 with the $d$ levels, responsible of the charge fluctuation.  $H_{dd}$, $H_{pp}$ and $H_{pd}$ are obtained from the DFT Hamiltonian using
 the wannierization procedure described above.

The electron-electron Coulomb repulsion in the $d$ shell reads: 
\begin{equation}
H_{{\rm Coul}}=\frac{1}{2}\sum_{m,m'\atop m'',m'''}
V_{mm''m'm'''}
\sum_{\sigma\sigma'}d_{m\sigma}^\dag d_{m''\sigma'}^\dag d_{m'''\sigma'}d_{m'\sigma},
\label{hcoul}
\end{equation}
For the evaluation of the Coulomb integrals $V_{mm''m'm'''}$ we transform the angular part 
to a basis of eigenstates of $\ell=2$. For the radial part we take 
an effective radial hydrogen-like function (
with  effective charge $Z$ and a  effective Bohr radius $a_\mu$) to
avoid the otherwise cumbersome numerical integration
of the actual Wannier functions.  In the basis of eigenstates of $\ell=2$,  all the Coulomb integrals  scale linearly with 
the value of $V_{0000}\equiv U$. \cite{Ferron15} 
Altought the numerical evaluation of $U$ in
terms of the parameters $Z$ and $a_\mu$  is straight-forward,
for the sake of generality, $U$ will be considered
a parameter, taken to satisfy 
the atomic  Hund's rule.    For a given choice of $U$,  $E_d$ is adjusted so that the average charge in the ground state of the many-body calculation described below equals the charge obtained in the DFT calculations.  

The spin-orbit term in the TM reads: 
\begin{equation}
H_{\rm SO}=\lambda_{SO}\,\sum_{mm',\sigma\sigma'} \langle m\sigma|\vec \ell\cdot \vec S|m'\sigma'\rangle
d_{m\sigma}^\dag d_{m'\sigma'},
\end{equation}
\noindent where $\lambda_{SO}$ is the atomic  spin-orbit coupling 
of the $d$-electrons.    This term is the only one that does not commute with the total spin operator and is the ultimate responsible of the lifting of the $2S+1$ degeneracy of the spin multiplets.

In the following we shall show results for the case of Fe on Cu$_2$N.  
Given the small size of the single particle basis (5 d orbitals and 6 p 
orbitals) and the fact that we restrict the Hilbert space to the configurations
$d^np^{12}$ and $d^{n+1} p^{11}$, with $n=6$ for Fe,  with a total of 1650 
multi-electron states, the multi-orbital Anderson model can be solved by exact 
numerical diagonalization.  

\subsection{Effective spin model \label{esm}}
 In the seminal work\cite{Hirjibehedin07} of Hirjibehedin {\em et al.}, 
the spin excitations measured with STM- IETS were found to be described with the following spin Hamiltonian: 
\begin{equation}
H= D\left(\vec{e}_1\cdot\vec{S}\right)^2  + E\left[\left(\vec{e}_2\cdot\vec{S}\right)^2 -\left(\vec{e}_3\cdot\vec{S}\right)^2\right]
\label{HHH}
\end{equation}
with $S=2$ and $\vec{e}_1=(0,1,0) $ along the Nitrogen direction, and $\vec{e}_2$ and $\vec{e}_3$ are the off-plane and the hollow directions. 
The experimental results could be fitted with $D=-1.55$meV and $E=0.31$ meV.   Thus, the wave function of the ground state and first excited state would be given by linear combinations of the rates of $|2,\pm 2\rangle$, with a small mixing with the state $|2,0\rangle$ in the case of the ground state.  The height of the inelastic steps was found to be related to the matrix elements of the spin operators, giving additional support to the notion that the quantized spin Hamiltonian (\ref{HHH}) provides a quite good description of the spin excitations of iron on this surface. 
Recent experiments\cite{Yan2015} with a detailed study of the IETS of single Fe/Cu$_2$N as a function of the three components of the magnetic field show that the addition of extra terms in the Hamiltonian (\ref{HHH}) yields an
even better agreement with the experiment. 
Spin chains formed with Fe atoms in this system can also be modeled successfully with  this Hamiltonian and the addition of interatomic Heisenberg coupling.
\cite{Bryant13,Spinelli14}  Altogether, these results support the notion that Fe can be described with a quantized anisotropic spin $S=2$.

\subsection{Adiabatic continuity and spin conservation argument \label{ac}}
We now address the crucial question: given that according to DFT, both charge and spin of the $d$ electrons in the transition metal are {\em not} quantized,  what is the origin of the quantized spin in the model Hamiltonian?.   We now show that the quantized spin belongs to the many-body wave function that combines configurations  with different charge states in the $d$ shell.  In the case of Fe,
these would be configurations  $d^6p^{12}$  with  $S^{(d)}=2$ and 
configurations with   $d^7p^{11}$ and $S^{(d)}=\frac{3}{2}$, where $S^{(d)}$ is the spin of the $d$ electrons.  These many-body states yield non-integer charge and magnetic moment in the TM, in agreement with DFT,    but they have a well defined {\em total} quantized spin, in agreement with the spin quantized models. 
This is strictly true     in the absence of spin-orbit interactions, and remains true when spin-orbit splittings are much smaller than the energy gap between different multiplets, which 
we show to be the case in a wide range of parameters.

In order to understand the results of our numerical calculations,
 it is convenient to remind that 
atomic iron, with 6 $d$ electrons, has a ground state with $S=2$ and $L=2$, and a total degeneracy of $(2L+1)(2S+1)=25$.
The former ground state is captured by
the multi-orbital Anderson model 
when  switch off  the $H_{dd}$ crystal field, the $H_{pd}$ hybridization, the SOC, considering configurations with  6 electrons. 
 The crystal field $H_{dd}$  quenches  the orbital moment, so that in the absence of  spin-orbit coupling, the ground state of the model has $S=2$ and no orbital degeneracy.  This multiplet is  separated from the next  higher energy multiplets, also with $S=2$,   by a gap of at least 300 meV, although this number depends on $U$ . 
 
\begin{figure}
\begin{center}
\vspace{15pt}
\includegraphics[width=0.4\textwidth]{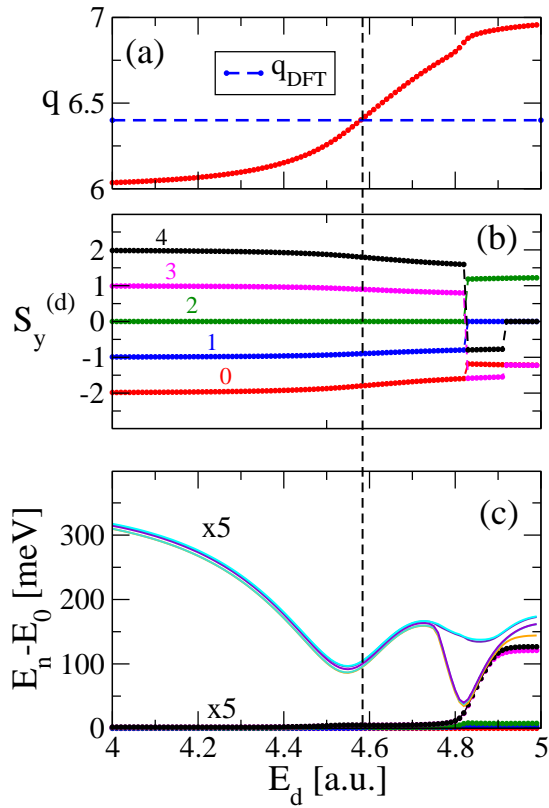}
\end{center}
\vspace{-15pt}
\caption{(Color online) Numerical diagonalization of the multi-orbital
Anderson model for Fe atom at Cu$_2$N.
(a) Average charge   of the $d$-levels of Fe atom for the Ground State
as a function of $E_d$  in atomic units ($U=5.5$ eV and $\lambda_{SO}=50$ meV).
(b) $S_y$ ($y$ is the N atoms axis) of the $d$ electrons for the $5$
lowest eigenvalues as a function of $E_d$ ($U=5.5$ eV and $\lambda_{SO}=0$).
(c) Excitation energies corresponding, for small $E_d$,  to the 2  lowest multiplets with $S=2$, as  a function
of $E_d$ ($U=5.5$ eV and $\lambda_{SO}=50$ meV) .}
\label{ci1}
\end{figure}

 We now discuss the effect of mixing configurations  with a different number of electrons in the $d$ shell. On one hand,
the relative weight of the configurations  
$d^6p^{12}$ and  $d^7p^{11}$ depends on $U$,
which is taken to be $U=5.5$eV. On the other hand,  
 $H_{pp}$, $H_{dd}$,  $H_{pd}$, are obtained from DFT  and $E_d$ 
remains as the only adjustable parameter in the calculation. We start with 
 a value of $E_d$ so that  the charge at Fe is  $q=6$ and ramp up $E_d$, so 
that $q$ increases, as shown in Fig. \ref{ci1}(a).  In Fig. \ref{ci1}(c) we 
plot the evolution of the excitation energies as a function of $E_d$.   For 
$q=6$ the  ground state has a multiplicity of 5. The next multiplet lies 300 
$meV$  above.  As we ramp up $q$,  these two multiplets remain well separated 
in energy,  quite beyond the point where $q=6.4$, the value obtained from DFT. 
We thus see that the ground state multiplet at  $q=6.4$ is {\em adiabatically 
connected}  to the ground state multiplet at $q=6$.    The effect of   
spin-orbit coupling, that we discuss in detail below, is to  create  a small 
splitting and to  mix  different states within the multiplet  as $E_d$ is 
varied, implying a change in the magnetic anisotropy tensor.  

The total spin is preserved as we ramp $E_d$.
However, by changing the 
relative weight of $d^6p^{12}$ and $d^7p^{11}$ configurations, 
 the  magnetic moment in the $d$ levels is expected to be reduced,  moving  
from $S(d)=2$ towards $S(d)=3/2$ . This 
is reflected in our calculations (see Fig. \ref{ci1}(b)), taking 
$\lambda_{SO}=0$  and a finite 
magnetic field $B_y$  that lifts the $2S+1$ degeneracy. As $E_d$ is 
increased,  the expectation value of  the operator describing the spin of the 
$d$ electrons along the $y$ axis,  $S^{(d)}_y$, calculated with  the 5  
lowest energy states,   evolves from the eigenstates of $S_y$ for $S=2$ to 
non-quantized values, in agreement with DFT results.

Further increase of $E_d$ yields that the mixing between $d^6p^{12}$ and 
$d^7p^{11}$  is so large that the splitting between the ground state $S=2$ 
multiplet and the first excited multiplet vanishes. When such regime is 
reached,  the spin of the ground state changes, 
breaking the adiabatic connection with the state with $S=2$ and quantized 
charge.  In our calculations this happens for $q\simeq 6.8$, larger than 
the DFT charge, $q>6.4$.  Thus, the model captures the crossover from the  
weak coupling limit,  where the ground state adiabatically connected with the 
$q=6, S=2$, state,  to the strong coupling limit in which the spin of the 
ground state multiplet changes  and the adiabatic connection is lost.

\subsection{TM-Nigrogen spin correlation \label{nsc}}

The fact that the quantized spin corresponds to configurations with two charge states involving both $d$ electrons in the TM and $p$ electrons in the first neighbor Nitrogen atoms has implications on the spin correlation of the TM and N magnetic moments.  Since  both  $d^n p^{12}$ and $d^{n+1} p^{11}$ have the same total spin,   we have $S(d^n)=S_T= S(d^{n+1})\pm \frac{1}{2}$.  The sign, and thereby the spin-correlation between the unpaired electron in the ligand and the magnetic moment of the atom,  depends on whether $S(d^n)$ is larger or smaller than $S(d^{n+1})$.  Thus,  for Fe we have
that $S(d^6)=2$ and $S(d^7)=3/2$,  so that the unpaired fermion must couple ferromagnetically with the $S=3/2$  of the $d^7$ configuration, to keep $S=2$.  In
contrast, for Cr we have $S(d^4)=2$ mixing with $S(d^5)=5/2$ configurations, to bring up the charge,  so that the unpaired fermion in the Nitrogen must couple antiferromagnetically with the atomic magnetic moment. 

This argument accounts for the trend obtained in our DFT calculations, shown in Fig. \ref{mn}, where the small magnetization of the first-neighbor nitrogen atoms is antiparallel to the TM magnetic moment for Ti, V, and Cr, and is ferromagnetic in the case of Fe, Co and Ni.  In the case of Mn we obtain both signs, depending on $U$.  From the argument of the previous paragraph we would expect a ferromagnetic coupling.
Incidentally, the same argument can be applied to the conventional  Anderson model with $S=1/2$,  predicting correctly the well known\cite{Schrieffer_old} antiferromagnetic interaction between the local moment and the adjacent electrons.

\subsection{Spin-orbit coupling, magnetic anisotropy and symmetry of the wave functions \label{swf}}
The discussion above has ignored the role of spin-orbit coupling, even if the numerical results shown in Fig. \ref{ci1}(a) and Fig. \ref{ci1}(c) are 
obtained with a spin orbit coupling $\lambda_{SO}=50$meV.  
These calculations show 
that the spin-orbit coupling splits the otherwise $(2S+1)$ degenerate
multiplets. However, the splittings
between the different multiplets are still much
larger than the energy gaps between them, so that a 
well defined total spin $S$ can be atributed, even in the presence of 
spin-orbit coupling.  The fine structure within the lowest energy multiplet 
can be described with an effective spin Hamiltonian, like the one in Eq.
(\ref{HHH}).

\begin{figure}
\begin{center}
\vspace{15pt}
\includegraphics[width=0.4\textwidth]{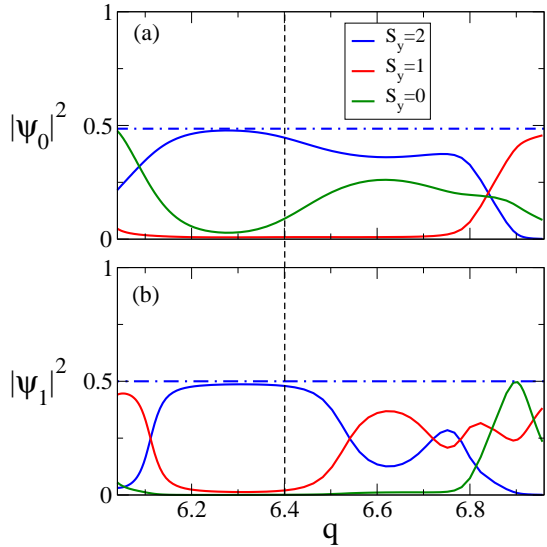}
\end{center}
\vspace{-15pt}
\caption{(Color online) Numerical diagonalization of the multi-orbital
Anderson model for Fe atom at Cu$_2$N ($U=5.5$ eV and $\lambda_{SO}=50$ meV).
Eigenvectors projection over $S_y$ for the ground state (upper panel)
and the first excited state (lower panel) as a function of the charge
in the $d$-shell.
Blue lines show projection over $|\pm2\rangle$, red lines show projection
over $|\pm1\rangle$ and green line shows projection over $|0\rangle$.
Blue dashed line shows the projection over $|\pm2\rangle$ obtained from the  
spin model. \cite{Hirjibehedin07}}
\label{ci2}
\end{figure}

\begin{figure}
\begin{center}
\vspace{15pt}
\includegraphics[width=0.4\textwidth]{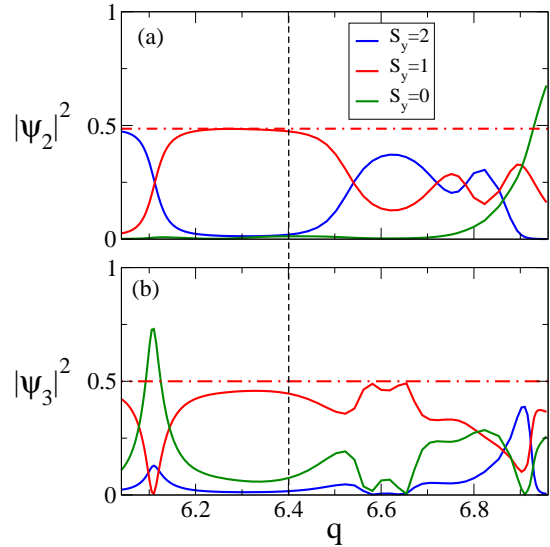}
\end{center}
\vspace{-15pt}
\caption{(Color online) Numerical diagonalization of the multi-orbital
Anderson model for Fe atom at Cu$_2$N ($U=5.5$ eV and $\lambda_{SO}=50$ meV).
Eigenvectors projection over $S_y$ for the second excited state (upper panel)
and the third excited state (lower panel) as a function of the charge
in the $d$-shell. Blue lines show projection over $|\pm2\rangle$, red lines
show projection over $|\pm1\rangle$ and green line shows projection over
$|0\rangle$.
Red dashed line shows the projection over $|\pm1\rangle$ obtained from the 
spin model. \cite{Hirjibehedin07}}
\label{ci3}
\end{figure}

\begin{figure}
\begin{center}
\vspace{15pt}
\includegraphics[width=0.4\textwidth]{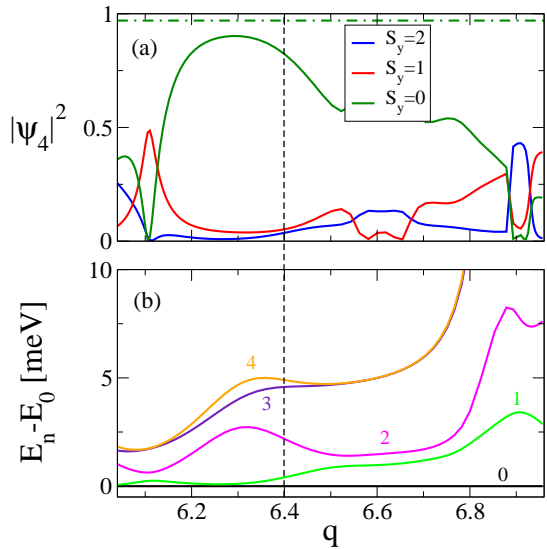}
\end{center}
\vspace{-15pt}
\caption{(Color online) Numerical diagonalization of the multi-orbital
Anderson model for Fe atom at Cu$_2$N ($U=5.5$ eV and $\lambda_{SO}=50$ meV).
(a) Eigenvectors projection over $S_y$ for the fourth excited state 
 as a function of the charge in the $d$-shell. Blue lines show projection 
over $|\pm2\rangle$, red lines show projection
over $|\pm1\rangle$ and green line shows projection over $|0\rangle$.
Green dashed line shows the projection over $|0\rangle$ obtained from the  
spin model. \cite{Hirjibehedin07}
(b) Low energy spectrum as a function of the charge in the $d$-shell.}
\label{ci4}
\end{figure}

We now discuss the symmetry of the wave functions obtained from the exact 
diagonalization for two values of $E_d$, corresponding to having either 
exactly $q=6$ electrons at the $d$ shell or $q= 6.4$, the average charge 
obtained from DFT. For that matter, we use the fact that the eigenstates 
$\psi_n$ of the multi-orbital Hamiltonian can be written as linear 
combinations of configuration states with well defined total $S_y$:
\begin{equation}
|\psi_n\rangle =\sum_{S_y,\gamma} \psi_n(\gamma,S_y) |\gamma,S_y\rangle
\end{equation}
where $\gamma$ labels all the other quantum number necessary to characterize 
the basis set.   In figures \ref{ci2}(a,b), \ref{ci3}(a,b) and \ref{ci4}(a) we 
plot the projection of the 5 lowest energy states of the multi-orbital 
Anderson model over the eigenstates of $|S=2, S_y\rangle$. 
In the case of $q=6$ it is apparent that the ground state 
wave function is dominated by $S_y=0$ states, in disagreement with the 
experiment (see Fig. \ref{ci2}(a)).  The arrangement of the energy levels 
seems to indicate that the Nitrogen direction ($y$ axis) is a hard axis in 
the problem ($\vec{e}_1=(0,1,0)$ and
$D>0$ in Eq. (\ref{HHH})). Given that the charge fluctuations are  negligible in this limit, 
the main contribution to the magnetic anisotropy comes from the interplay 
between the crystal field term in the Hamiltonian, $H_{dd}$ and the spin-orbit 
coupling. Interestingly, when $E_d$ is ramped so that $q$ increases, the 
content of the wave functions evolves and for $q=6.4$ the wave functions 
(Fig. \ref{ci2}, Fig. \ref{ci3} and Fig. \ref{ci4}(a)) are in good agreement 
with those obtained from the spin model\cite{Hirjibehedin07} in which the 
Nitrogen direction is the easy axis in the problem ($\vec{e}_1=(0,1,0)$ and 
$D<0$ in Eq. (\ref{HHH})).  In particular, we note 
that the five lowest energy states of the Anderson model have wave-functions 
with strong overlap with those of the effective spin 
Hamiltonian.\cite{Hirjibehedin07} 
We thus see that the ligand field contribution, coming from the $dp$ 
hybridization,  changes qualitatively the magnetic anisotropy tensor. Interestingly,  in the 
case of Fe/Cu$_2$N we find that  the inclusion of charge fluctuations is 
essential to capture the correct easy axis within the multi-orbital Anderson 
model.

We finally analyze the low energy excitation spectrum, $E_n-E_0$, where $E_0$ 
is the energy of the ground state.  In the range of $E_d$ considered, such 
that $q$ moves from the nominal value  $q=6$ to the DFT value $q= 6.4$,  
the five energy levels of the lowest energy multiplet are always split. This 
is expected in the case of a integer spin ($S=2$) described with Hamiltonian 
Eq. (\ref{HHH}).  Interestingly, the  low energy splittings increase as $q$ is 
increased towards the DFT value, as shown in Fig. \ref{ci4}(b).   
This is related
to the fact that the energy gap between the first and second five-fold 
degenerate multiplet decreases, as shown  in Fig. \ref{ci1}(c).
This behavior can be 
understood in terms of degenerate perturbation theory,
where spin-orbit 
coupling only splits the states in the lowest energy multiplet through virtual 
transitions to the  higher state multiplets.  
We note that the spin excitation 
energies obtained from our calculation are a $25$ percent smaller than those 
observed in the experiment.  So, whereas the model captures the right 
symmetry, it can only give a rough description of the excitation energies, 
which is probably due to the approximations in the model, 
such as the  restrictions taken in the many-body Hilbert space.

\section{Conclusions \label{c}}

In this work we have undertaken a systematic study of the electronic properties of the 3d transition metals on the Cu$_2$N surface. 
We systematically find that the charge and spin of the $d$ electrons are not 
quantized, and are thereby  different from the one in isolated atoms, which is 
expected given the conducting nature of the substrate. We have then addressed 
the issue of how to reconcile these results with the fact that quantized spin 
models account for the spin excitations of Mn, Fe and Co ad-atoms.   For that 
matter we propose a multi-orbital Anderson model in which many-body states 
that mixes configurations with two charge states in the $d$ shell are 
considered.  Importantly, even if the charge is not well defined in the $d$ 
shell, these multi-electron wave functions have a well defined total spin 
$S$.  We find that the states with a quantized charge in the $d$ shell are 
adiabatically connected with the actual many-body states that mix 
configurations $d^n p^{12}$ and $d^{n+1}p^{11}$ in the sense that both have 
the same 
total spin $S$ and there is no mixing with higher energy multiplets as the 
addition energy is varied numerically.  

 We thus conclude that quantized spin 
$S$ of the model actually refers to the spin of these many-body states that 
include both the $d$ electrons and the ligand electrons.    It is thus fair to 
say that the magnetism in this system is not strictly atomic, which connects  
with previous results in the case of magnetic atoms 
on metallic surfaces, \cite{Lounis11} for which sophisticated theoretical 
treatments have been proposed .\cite{Lounis14,Lounis15}   
   Our  picture provides a natural explanation to the 
sign of the spin correlation between the TM and the nitrogen atoms obtained in 
the DFT calculations and reconciles the use of quantized spin Hamiltonians 
with the results of DFT calculations. Our Anderson model  calculations for 
Fe/Cu$_2$N also indicate that a charge fluctuations in the Fe $d$ shell, 
due to the hybridization to the ligands, are essential to capture both the 
symmetry and the magnitude of the magnetic anisotropy observed in the 
experiment. 

Finally, we expect that  our analysis should also be applicable to other systems with magnetic adatoms  and molecules deposited in conducting surfaces
whose spin excitations can be described in terms of quantized spin models\cite{Lounis11,FePc,Khajetoorians10,Khajetoorians13,Jacobson15}

\section*{Acknowledgments}

JFR acknowledges  financial supported by MEC-Spain (FIS2013-47328-C2-2-P) 
and Generalitat Valenciana (ACOMP/2010/070), Prometeo. This work has been 
financially supported in part by FEDER funds. JLL and JFR acknowledge financial
support by Marie-Curie-ITN 607904-SPINOGRAPH. JLL and AF thank the hospitality 
of the Departamento de F\'{\i}sica Aplicada at the Universidad de Alicante.
AF acknowledges funding from the European Union's Seventh Framework Programme 
for research, technological development and demonstration, under the PEOPLE 
programme, Marie Curie COFUND Actions, grant agreement number 600375 and 
CONICET. We thank F. Delgado for fruitful discussions.

\end{document}